# The Size of the Emitting Region in the Magnetic Eclipsing Cataclysmic Variable Stars

Kateryna D. Andrych[1], Ivan L. Andronov[2],
[1] Department of Astronomy, Odessa National University
[2] Department "High and Applied Mathematics"
Odessa National Maritime University

**Abstract.**
We discuss a method for determination of the size of the emitting region close to the compact star in a binary system with eclipses by a secondary, which fills its Roche lobe. The often used approach is to model the Roche lobe by a sphere with the "effective radius" corresponding to the volume of the Roche lobe. This approach leads to a 4-6% overestimate of the radius, if taking into account the angular dimensions of the Roche lobe seen form the compact star. Andronov (1992) had shown that the projection of the Roche lobe onto the celestial sphere is close to an ellipse and had tabulated these dimensions as a function of the mass ratio. Also he published the coefficients of the approximation similar to that of the Eggleton (1983) for the "sphere corresponding to the same volume". We compare results obtained for the "circle+circle", "ellipse+circle" and "ellipse+point" approximations of the projections of the red dwarf and a white dwarf, respectively. Results are applied to the recently discovered eclipsing polar CSS 081231:071126+440405.

**Introduction**

Cataclysmic variables are excellent laboratories to study numerous processes taking place in the Universe. Especially interesting are the systems, in which the magnetic field of the white dwarf is strong enough to channelize the plasma flow (accretion stream) at far distances, namely, closer to the inner Lagrangian point. In this case, the spin and orbital periods of the magnetic white dwarf coincide, as was initially foung by S.Tapia (1977) for the prototype star AM Her. Because of the polarization of the emission, such objects are called "polars". Recent reviews on our results were published by Andronov (2007, 2008).

The object CSS 081231:071126+440405 (also called OTJ0711) was discovered by Denisenko and Korotkiy (2009) on the New Year night 31.12.2008, and was included in our list for monitoring according to the "Inter-Longitude Astronomy" campaign (Andronov et al. 2010). The observations were obtained during dozens of nights using CCD by Arto Oksanen (Finland), Pavol Dubovsky (Slovakia), Jon-Na Yoon et al (Korea). But only fast photometry with 1s resolution obtained by Sergej Kolesnikov using the 2.6m Shain telescope of the Crimean astrophysical observatory allowed determination of the (very short) duration of the ascending and descending branches of only 4.752±0.306 seconds.

Andronov and Andrych (2014) presented an analysis of the data assuming the mass ratio of components obtained using statistical relationship. In this paper, we discuss dependence of the angular dimensions of the Roche lobe on the mass ratio.

**Angular Dimensions of the Roche lobe**

The Roche lobe has a distorted shape, which prevents determination of the characteristics analytically. We need to apply a numerical analysis and determine the angles. They are listed in Andronov (1992). The distance from the center of the star to the Roche lobe is dependent on the direction, but, for many applications, the Roche lobe was approximated in the simplest way – as the sphere. As to the "effective radius", there are some approaches. The main one is to use the radius of "equal volume" $r_V$, i.e. the radius of the sphere, the volume of which is equal to the volume of the Roche lobe. We will use this index "V" ("volume"), despite in other papers it may be "e" ("effective") or "L" ("lobe"). Kopal (1959) tabulated these values, and Paczynski (1971) proposed a two-interval approximation:

$$r_V = \begin{cases} 0.46224\mu^{1/3} & \text{if } q < 0.523 \\ 0.38 + 0.2\lg q & \text{if } 0.523 \leq q \leq 20 \end{cases} \quad (1)$$

Here $q = M_2/M_1$ is called the "mass ratio" of two stars, with the indexes "1" and "2" corresponding to the primary and secondary (the Roche lobe of which is studied). The parameter $\mu = M_2/(M_1 + M_2) = q/(1+q)$. Although this approximation has a relative accuracy of <2%, the derivative of this function has a discontinuity of ~20%, what has no physical sense and made evolutionary computations less accurate.

Eggleton (1983) made numerical computations of the volume and proposed an analytical approximation

$$r_V = \frac{Aq^{2/3}}{Bq^{2/3} + \ln(1+q^{1/3})} \quad (2)$$

with coefficients $A$=0.49, $B$=0.6 and relative accuracy of 1% both for the function and its derivative. The asymptotic simplifications are $r_V \approx Aq^{1/3}$ for $q \ll 1$ and $r_V \approx A/B$ for $q \gg 1$. This approximation is extremely popular and the number of citations in the ADS approaches 1200.

Andronov (1982) preferred to use another characteristic of the Roche lobe, which is the "barotropic" radius $r_B$, which was introduced by Kopal (1959) as a characteristic radius of the Roche lobe. This the radius of the spherically symmetrical star with the same difference of the potentials between the center and the atmosphere as in the Roche lobe. This seems to be more reasonable approximation for stars with outer layers distorted by tidal forces. Detailed study of the distribution of density and pressure for polytropic models was presented by Sirotkin (1997). The corresponding asymptotic approximation

$$r_B = B\mu^{1/3} \quad (3)$$

was corrected as

$$\lg B = -0.3390 + 0.0561 \lg q + 0.0354 (\lg q)^2 \quad (4)$$

Such two-step approximation produces relative accuracy of 0.12% for $0.1 \leq q \leq 1$ and 0.5% for $0.07 \leq q \leq 1.5$.

To determine angular dimensions of the Roche lobe, Andronov (1992) determined directions of tangent lines to the Roche lobe. The tables were fitted by a function in form of Eggleton (1983). The main angles are in the orbital plane ($\sin\vartheta_0$, $A=0.449$ (a misprint 0.499 in the paper), $B=0.5053$, $\psi=0°$), perpendicular to the orbital plane ($\sin\vartheta_{90}$, $A=0.4394$, $B=0.5333$, $\psi=90°$), "effective" in a sense of equal solid angles of the Roche lobe and the sphere ($\sin\vartheta_e$, $A=0.4441$, $B=0.5195$).

For a comparison of these characteristic dimensions, let's list them for a sample value $q=1$: $r_V = 0.3799$ (Eggleton 1983), $r_B = 0.3636$, $\sin\vartheta_0 = 0.3746$, $\sin\vartheta_{90} = 0.3591$, $\sin\vartheta_e = 0.3666$, so we can see a significant difference (Fig.1).

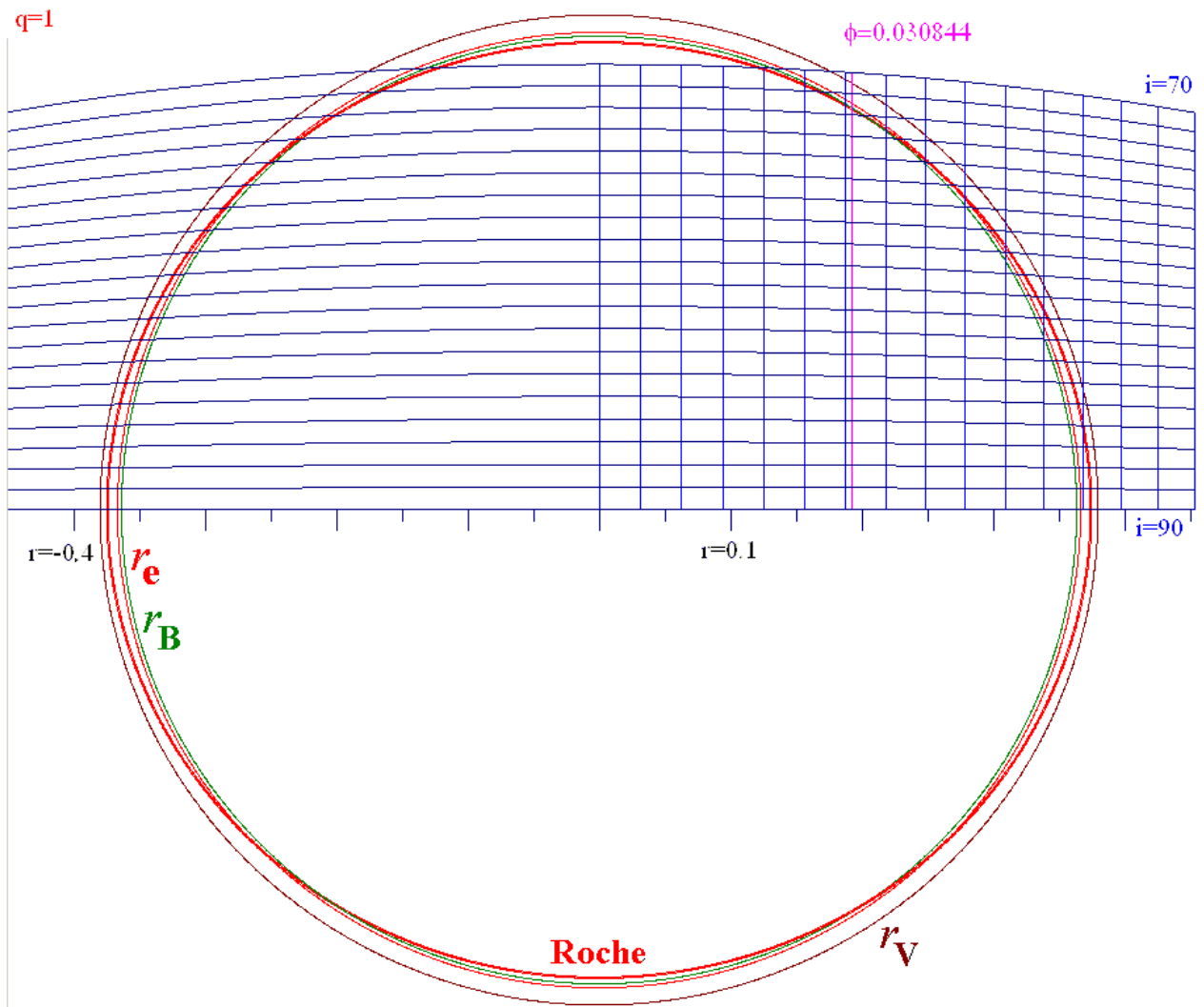

Figure 1. Different models of the Roche lobe seen from the center of the compact object: spheres of "effective" radiuses $r_B$, $r_V$, $r_e$ and the accurate nearly elliptic "Roche". ). Elliptic arcs correspond to the tracks of the observer (in the system of co-ordinates rotating with a binary system) for inclination $i$ from 70° to 90° (step 1°), Vertical lines correspond to phases of eclipse (step 0.005) (blue), and the violet line corresponds to the observed value $\phi = 0.030844$ for the object OTJ0711. The diagram is calibrated ($r$) in units of orbital separation $a$.

The true shape of the Roche lobe is marked by a thick red line "Roche". Even in the orbital plane, the largest visible dimension of the Roche lobe does not reach the value of $r_V$. The barotropic radius $r_B$ is much closer to the "effective" $r_e$, but, anyway, taking into account the elliptic shape will produce much more accurate results. The ratio $\sin\vartheta_0 / \sin\vartheta_{90}$ increases from 1.038 for small $q$ to 1.043 at $q=1$ and even to 1.093 at $q=19$.

In Fig.1. one may see that the use of $r_V$ instead of correct computations of $\sin\vartheta_{90}$ changes the limiting inclination by 1.3°. It will also cause different impact angles of the trajectory into the limb, so the estimates of the normal velocities, and so estimates of the size of the white dwarf from the duration of the ascending/descending branch.

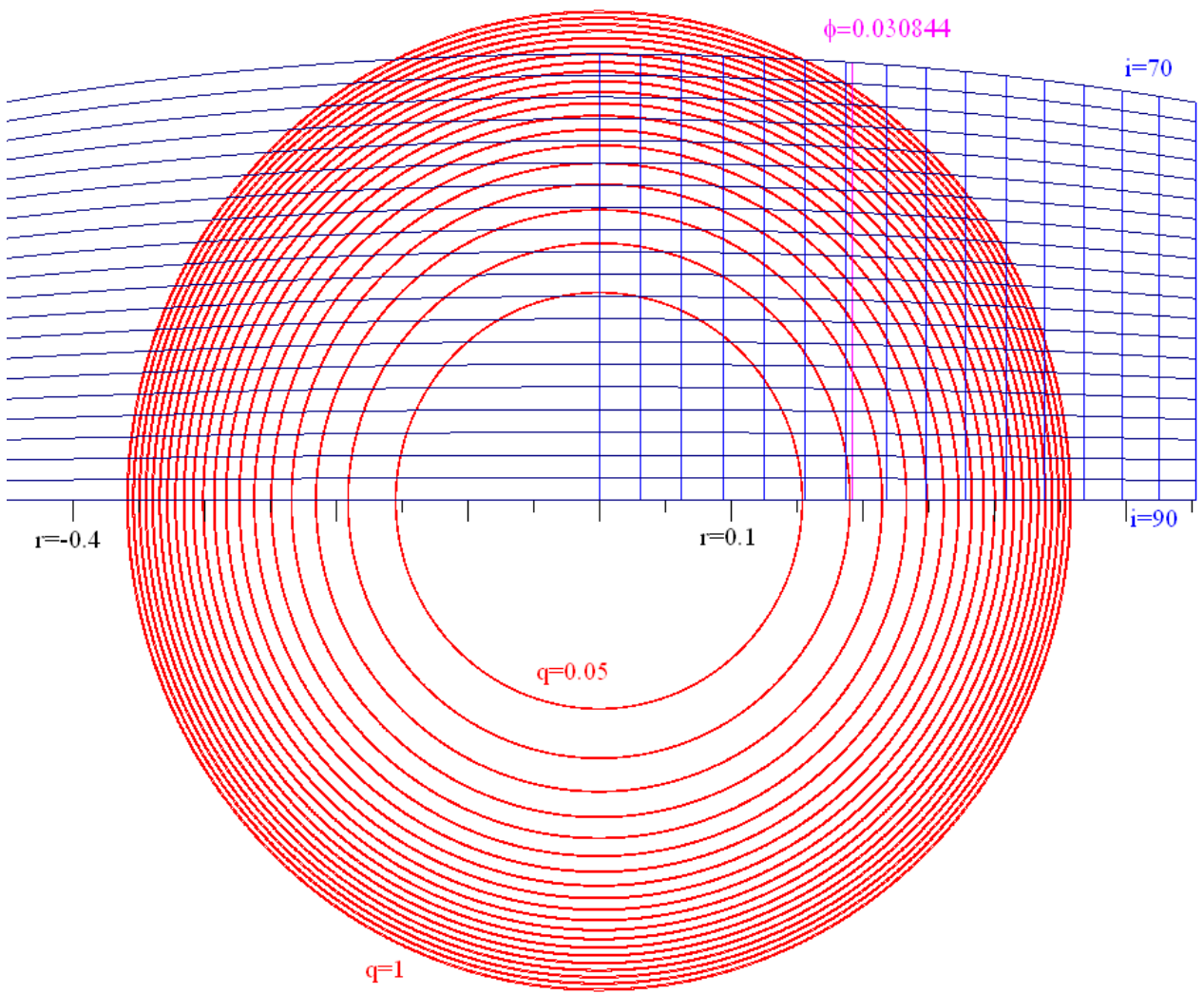

Figure 2. Nearly elliptic projections of the Roche lobe seen from the center of the white dwarf for the mass ratio $q$ from 0.05 to 1 (step 0.05) (red). Elliptic arcs correspond to the tracks of the observer (in the system of co-ordinates rotating with a binary system) for inclination $i$ from 70° to 90° (step 1°), Vertical lines correspond to phases of eclipse (step 0.005) (blue), and the violet line corresponds to the observed value $\phi = 0.030844$ for the object OTJ0711. The diagram is calibrated ($r$) in units of orbital separation $a$.

In Fig. 2, we show the shapes of the Roche lobe as seen from the center of the compact object for a set of different values of the parameter *m*. From the duration of the eclipse (Andronov and Andrych 2014), one may set also the minimal limit $\mu > 0.1$. They adopted the mass ratio of *q*=0.3 in this system and masses of $M_1$=0.543$M_\odot$ and $M_2$=0.163$M_\odot$. Then the parameters are the following: $r_V$=0.28103, $\sin\vartheta_0 = 0.27216$, $\sin\vartheta_{90} = 0.26219$. Using the observed duration of the eclipse $\phi$, one may determine two values for inclination of the orbit *i* for the elliptical (*i*=79.1177±0.0075°) and the spherical ($r_V$) approximations (*i*=77.1231±0.0029°). The difference of 2° shows a necessity of correct computations of the shape of the Roche lobe instead of old spherical approximation with the radius $r_V$.

The radius of the white dwarf may be determined from the "mass-radius" relation (Andronov and Yavorskij 1990)

$$\frac{R_{WD}}{R_\oplus} = 0.011153\left(\left(\frac{M_*}{M}\right)^p - \left(\frac{M}{M_*}\right)^p\right)^q$$

with *p*=2/3 and *q*=0.465. Here we corrected the misprint in the coefficient 0.01153 published in the paper by Andronov and Yavorskij (1990). The radius of the white dwarf is much smaller than that of the white dwarf, so the corrections to the duration of the descending/ascending branch of the light curve due to the non-linearity of the visible trajectory are negligible at a scale of the dimension of the white dwarf. If needed, these corrections may be computed as described by Andronov and Andrych (2014). The size of the hot spot is 1300km. This is much smaller than the size of the white dwarf, so this "hot spot" may be the accretion column impacting the atmosphere of the white dwarf.

*Acknowledgements.* This study is a part of the projects "Inter-Longitude Astronomy" (Andronov et al., 2010, 2014) and "Ukrainian Virtual Observatory" (Vavilova et al., 2012). We thank Dr. Bogdan Wszołek for excellent hospitality.